\documentclass[prb,twocolumn,amsfonts,showkeys,footinbib,superscriptaddress,byrevtex,citeautoscript]{revtex4-1}

\usepackage[dvips]{graphicx}
\usepackage{dcolumn}
\usepackage{amsmath}

\newcommand{\defe}[2]{#1$_{\mathrm{#2}}$}
\newcommand{\defeq}[3]{#1$_{\mathrm{#2}}^{#3}$}
\newcommand{\Cd}[0]{\mathrm{Cd}}
\newcommand{\Te}[0]{\mathrm{Te}}
\newcommand{\V}[0]{\mathrm{V}}

\begin{document}
\title{Tellurium vacancy in cadmium telluride revisited: size effects in the electronic properties}
\author{E. \surname{Men\'{e}ndez-Proupin}}
\affiliation{Departamento de F\'isica, Facultad de Ciencias, 
Universidad de Chile, Las Palmeras 3425, 780-0003 \~Nu\~noa, Santiago, Chile}
\author{W. \surname{Orellana}}
\affiliation{Departamento de Ciencias F\'isicas,  
Universidad Andres Bello, Rep\'ublica  220, 037-0134 Santiago, Chile}
\date{\today}
\begin{abstract}
The quantum states and thermodynamical properties of the Te vacancy in CdTe are addressed by first principles calculations, including the supercell size and quasiparticle corrections. It is shown that the 64-atoms supercell 
calculation is not suitable to model the band structure of the isolated Te vacancy. This problem can be solved with 
a larger 216-atoms supercell, where the band structure of the defect seems to be a perturbation of that of the perfect crystal. It is interesting to note that the Te-vacancy formation energy calculated with both supercell sizes are close in 
energy, which is attributed to error cancelation. We also show  that the interplay between supercell size effects
and the band gap underestimation of the generalized gradient approximation strongly influences the predicted 
symmetry of some charge states.
\end{abstract}

\maketitle

\section{Introduction}

Cadmium telluride (CdTe) is a semiconductor crystal employed as photovoltaic material which 
is growing rapidly in acceptance, becoming a cheap alternative to the silicon solar cells. CdTe 
has advantages for photovoltaic applications, such as a direct band gap, allowing to absorb solar 
energy in thin films. Its band gap of 1.5 eV at room temperature is in the range for maximum 
conversion efficiency of a single absorber solar cell\cite{shockley}. CdTe is also widely used for 
radiation detectors\cite{librocdte}. However, the nature of the defects in CdTe thin films has been subject of 
controversy for long time, and its identification frequently relies on thermodynamics arguments 
and ab initio calculation of the defects and impurities formation energies. 

Electron paramagnetic resonance (EPR) spectra unveiled a paramagnetic center that was initially 
assigned to the tellurium vacancy \defe{V}{Te}\cite{vte_epr}. Initial evidences suggested that 
its charge states are neutral and singly positive, with the transition level 
$\varepsilon(+/0)=0.2$~eV\cite{vte_epr}. However, this assignment has been rejected based on both 
density functional theory (DFT) calculations of total energies\cite{lany2001vte,vcd_hse} and 
hyperfine interaction terms\cite{illgner96vte}. Later on, this EPR signal was reassigned to the 
\defe{Te}{Cd} antisite\cite{verstraten2003}. No direct evidence of \defe{V}{Te} has been obtained 
afterwards from experiments, but from theoretical calculations this defect is one of the most important 
intrinsic donors.\cite{dusingh08,du2012,weiprl2013,wei2014}

The formation energy of  \defe{V}{Te} can be obtained computing the difference between the total energies of a crystal supercell containing  $n$ CdTe units,  $E(\Cd_{n}\Te_{n})$,  and the energy of the same supercell with a missing Te, $E(\Cd_{n}\Te_{n-1})$, supplemented with the Te chemical potential $E(\Te)+\Delta\mu_{\Te}$
\begin{eqnarray}
\Delta  H_f ({\V_{\Te}}^q &&) =  E(\Cd_{n}\Te_{n-1}) +E(\Te) - E(\Cd_{n}\Te_{n})    \nonumber  
 \\
+&&  \Delta\mu_{\Te}+ q [E_V + E_F]  + \Delta  E_{size} + \Delta E_{q.p.}.
\label{eq:one}
\end{eqnarray}
In Eq.~\ref{eq:one}, $E(\Te)$ is the energy of Te in a reference state, in this case it is the energy per atom of  bulk Te. $\Delta\mu_{\Te}$ is the chemical potential relative to the reference state, and it depends on the thermodynamic equilibrium conditions. In Te-rich conditions $\Delta\mu_{\Te}=0$, while in Te-poor conditions $\Delta\mu_{\Te}$ equals the CdTe formation energy\cite{zhangwei2002}.  For a charged vacancy, $q$ is the number of electrons donated to the environment, and $E_V + E_F$ is the electron chemical potential expressed as the sum of the valence band maximum $E_V$ and the Fermi level $E_F$.
 $\Delta  E_{size}$ and $\Delta E_{q.p.}$ are size and quasiparticle corrections that are explained in appendices \ref{app:size} and   \ref{app:qp}, respectively.

Most calculations of CdTe point defects have been investigated using the 64-atom supercell 
(SC64) of the perfect crystal.\cite{zhangwei2002,dusingh08,du2012,weiprl2013,wei2014,vcd_hse,cdte_ggau} This 
supercell is $2\times 2\times 2$ times the 8-atoms cubic unit cell. In this article, we 
show that the SC64 supercell is too small to obtain reliable results for the \defe{V}{Te}
band structure, while thermodynamic calculations are hampered by  uncontrollable size effects 
that are incompatible with the notion of diluted defects. We observe that changes in the conduction 
band are so dramatic that destroy the reference to apply quasiparticle gap corrections to the formation 
energy of negatively charged states. We show that this problem is solved using a 216-atom
supercell of the perfect crystal (SC216), which is $3\times 3\times 3$ times the 8-atom cubic 
unit cell. In addition, the band filling induced by the use of a k-point grid causes a 
spurious distortion in the double negatively charged state $2-$. We show that this size effect 
can be avoided by analyzing the band diagram and applying the band-filling correction. 
According to our results, 
a large supercell with only one k-point is enough to avoid this problem. For CdTe, this 
large supercell can be $4\times 4\times 4$ times the cubic unit cell, that contains 512 atoms for 
the perfect crystal (SC512).

\section{Computational details}
\label{sec:methods}

Our DFT calculations were performed using a plane-wave projector augmented wave\cite{paw1,paw2} 
scheme, as implemented in the Vienna Ab Initio Simulation Package (VASP)\cite{vasp4}. We have 
used mostly the generalized gradient approximation (GGA) for the exchange-correlation functional 
as proposed by Perdew, Burke, and Ernzerhof (PBE)\cite{pbe}. In certain cases, we have also used 
the hybrid functional proposed by Heyd, Scuseria and Ernzerhof (HSE06)\cite{hse06}. This 
functional generally allows one to obtain better band gap energies and structural properties than 
those obtained with PBE, but at the cost of a great increment in computer time.  

We have used a plane-wave cutoff energy of 460 eV for the wavefunctions. To sample the reciprocal space, 
we have used  $\Gamma$-centered $3\times 3\times 3$ and $2\times 2\times 2$  k-point grids for
the SC64 and SC216 supercells, respectively. This setup allows convergences of 0.5 meV/atom and 
0.4 kbar in total energy and pressure, respectively, as tested for variable cell relaxations of 
the CdTe zincblende unit cell with $8\times 8 \times 8$ k-point grid. In fact, a reduced cutoff 
of 285 eV allows a convergence within 10 meV/atom in total energy and 2 kbar in pressure, but the 
properties derived from relative energies, e.g., lattice constants and bulk moduli are well 
converged. The CdTe lattice constant obtained with variable cell relaxation with a cutoff of 285 eV 
differs by 0.4 \% from the converged value, but when the energy is fitted with the Birch-Murnaghan
equation of state, the difference falls to 0.005 \%. The energy difference between the zincblende and 
wurtzite structures is 3.5 meV/atom considering both cutoffs.
The $n\times n\times n$ supercell size is $n$ times the converged theoretical 
lattice constant of the perfect crystal (6.6212 \AA). 
 We have used a  cutoff of 285 eV to relax the 
point defect structures with fixed supercell size, and a cutoff of 460 eV to compute the total energies. 
The highest precision is needed because the energy error is amplified by the number of atoms in the 
supercell, e.g., for the neutral Cd vacancy in SC64 supercell, the formation energy computed with 
both cutoffs differs by 0.03 eV. 
We have verified that for higher cutoff energy, 
the forces on the atoms in the relaxed geometry keep below 0.01 eV/\AA.

 In previous works, the
model SC64 has generally been used with $2\times 2\times 2$ $\Gamma$-centered k-point 
grids.\cite{zhangwei2002,dusingh08,du2012,weiprl2013,wei2014} This sampling is equivalent to using only the $\Gamma$-point  
for the SC512 supercell. 
For \defeq{V}{Te}{2+}, using the $2\times 2\times 2$ and $3\times 3\times 3$ grids,  the 
difference in formation energy is 0.030~eV for model SC64. For model SC512 with $\Gamma$-point sampling,  the error in total energy is scaled to 0.24~eV.

The calculated CdTe heat of formation is found to be of 0.93 eV, in good agreement with the available
experimental value of 0.96 eV\cite{du2012}. The energy of bulk Cd has been obtained by relaxing the 
2-atom unit cell of the experimental structure ICSD \#52793 available from the MedeA$^{\circledR}$ software 
environment.\cite{medea} The Brillouin zone was sampled by a $22\times 22\times 10$ k-point grid.  The energy of bulk Te has been obtained relaxing 
the crystallographic 3-atoms unit cell ICSD \#23060. The Brillouin zone was sampled by a $8\times 8\times 5$ 
k-point grid and tested with a $15\times 15\times 10$  grid. 

The formation energies have been corrected for the effects of supercell size and quasiparticle corrections 
to the band edges, as described in the Appendices.

\section{Results and discussions}
\subsection{The Te vacancy in a 64-atom supercell: Structural and electronic properties}
\label{sec:iiia}
The SC64 supercell is the standard framework nowadays to model defect in diamond- or zincblende-type 
semiconductors. It is a popular choice because of its cubic shape, that maximizes the distance between 
the defect and its periodic images, and allows easy charge corrections. In addition, it is affordable for 
high level methods beyond GGA of DFT. The next cubic supercell, containing 216 atoms, is still 
extremely costly for high level methods. Next we analyze the band structure of the defective supercells, 
and show the problems to apply band-filling and quasiparticle corrections with the SC64 model. 

\begin{figure*}[htb!]
\begin{center}     
\includegraphics[width=.90\textwidth]{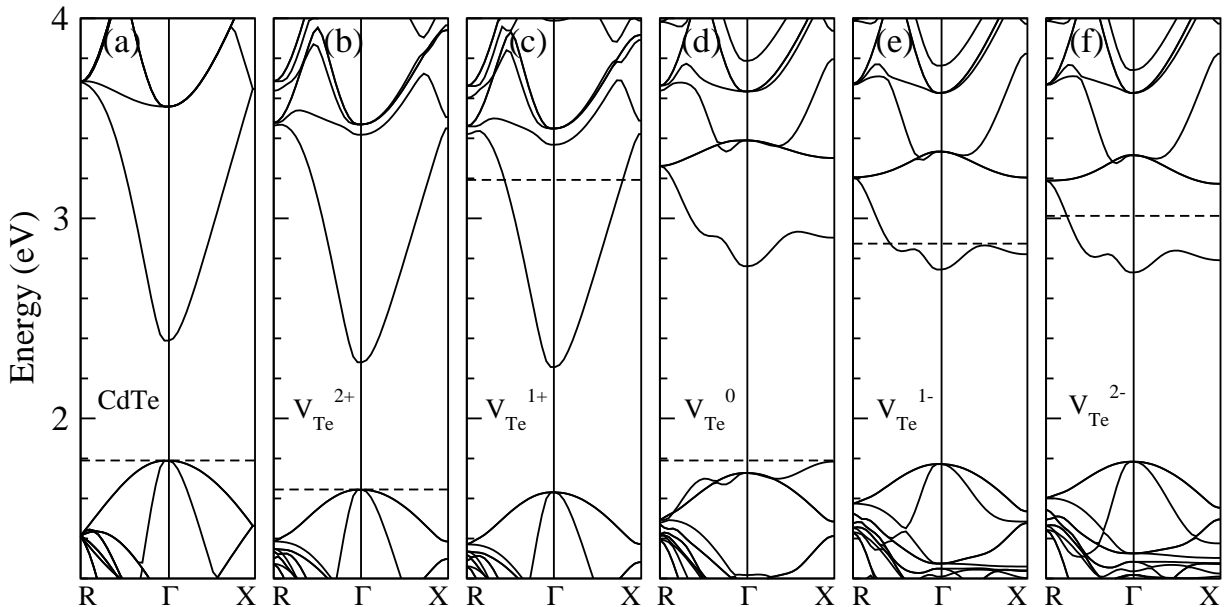} 
\caption{Band structure calculations for the Te vacancy in CdTe performed with a 64-atom 
supercell. (a) Perfect crystal, (b) to (f) Te vacancy in the charge states $2+$ to $2-$.
The dashed line indicates the highest occupied level.}
\label{fig:bandas64}
\end{center}
\end{figure*}

Figure \ref{fig:bandas64} shows the band structure calculations of the perfect CdTe and  \defe{V}{Te} 
in the SC64 supercell. The atomic  positions were relaxed keeping the $T_d$ symmetry, 
while distorted configurations will be considered separately using larger supercells.  The vacancy in the SC64 supercell is strictly a periodic model of \defe{V}{Te} in a concentration 
of one every 64 atoms or $10^{20}$~cm$^{-3}$.  
 In Figure~\ref{fig:bandas64} we see that
for the charge states $0,1-$, and $2-$, the conduction band (CB) is dramatically modified with a
significative increment of the band gap. This global change produced by the vacancy 
is incompatible with the notion that an impurity should introduce small perturbations on the 
 crystal 
band structure. This example shows that band diagrams, which are in principle not adequate 
to study aperiodic systems, are still useful to uncover flaws in defect models. 
From the practical side of thermodynamic calculations, the drastic alteration of the 
conduction band makes impossible to apply gap corrections, resulting in a false stability of 
the negatively charged states in n-type doping conditions. The only solution for the exaggerated  
influence of  \defe{V}{Te}  is to explore configurations with a larger supercell. 

\subsection{The Te vacancy in the 216-atom supercell: Structural and electronic properties}

\begin{figure*}[htbp]
\begin{center}
\includegraphics[width=.90\textwidth]{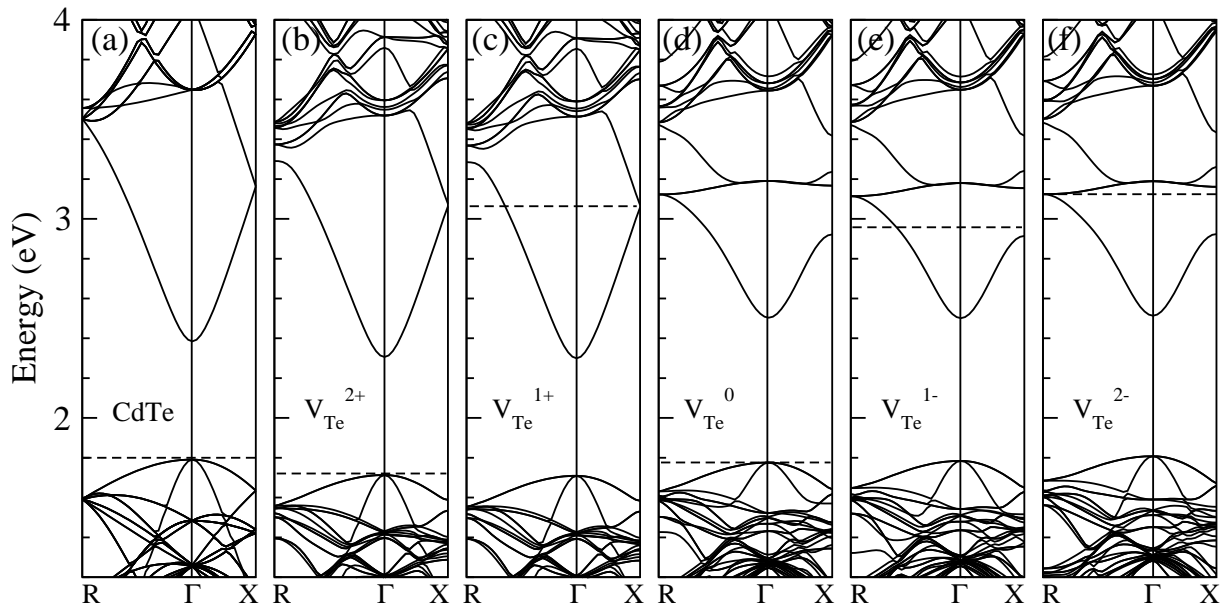}
\caption{Band structure calculations for the Te vacancy in CdTe performed with a 216-atom 
supercell. (a) Perfect crystal, (b) to (f) Te vacancy in the charge states $2+$ to $2-$. 
The dashed line indicates the highest occupied level.}
\label{fig:bandas216}
\end{center}
\end{figure*}

Figure \ref{fig:bandas216} shows the band diagram of the perfect crystal and the V$_{\rm Te}$
defect in different charge states computed with the $3\times 3\times 3$ supercell of the conventional 
8-atoms unit cell (SC216). The $T_d$ symmetry of the atomic arrangement around the vacancy has been conserved in the relaxation as for SC64. Distorted configurations with lower energy will be studied separately below. 
For all the charge states considered, the effects of 
the vacancy are small band shifts and splits. 
 In addition, in Fig.~\ref{fig:bandas216} we can recognize  an empty-defect band over 
3.2 eV for $q=2-,1-, 0$. As we will see later, the defect band becomes active,  driving the 
so called Jahn-Teller distortions. The origin of the defect band can be understood as follows: 
Each Te-Cd covalent bond is populated by 0.5 Cd electron and 1.5 Te electron. When \defe{V}{Te} 
is formed, four $sp^3$ bonding orbitals and six electrons are removed. This process leaves two  
electrons from the four surrounding Cd 5s orbitals. These orbitals combine to form four defect 
bands: one mixed with the host valence band (VB) that gets occupied by the electrons left by 
the Cd broken bonds, and three empty isolated bands ($\sim 3.1-3.5$ eV). Of these three bands, 
the lowest two are degenerate. In states $1-$ and $2-$, the defect levels keep empty and the 
extra electrons occupy perturbed host states (PHS),\cite{PHS} that in the supercell calculation 
appear similar to the CB.

\begin{figure}[tbp]
\begin{center}
\includegraphics[width=.48\textwidth]{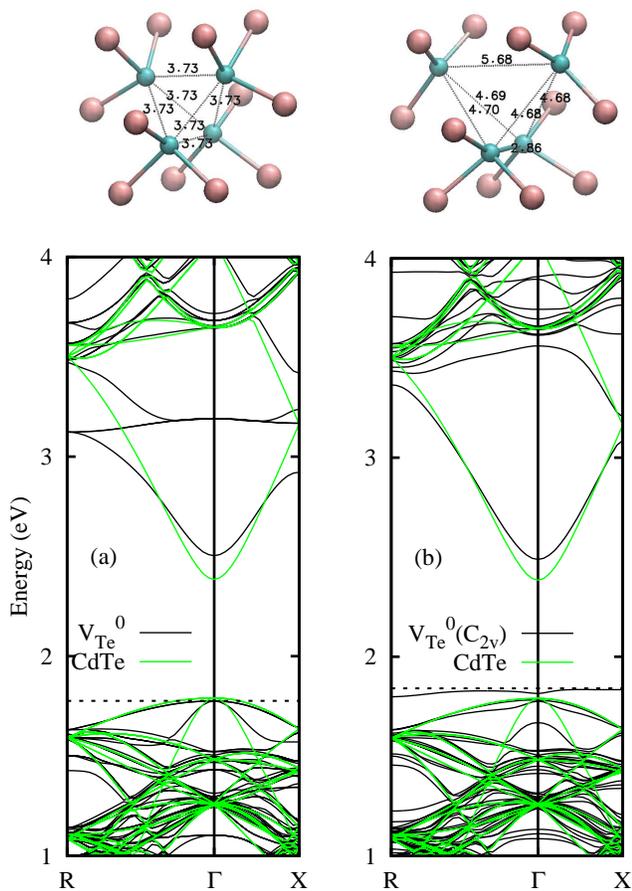}       
\caption{(Color online) Stable geometries and band diagrams for the neutral Te vacancy without (a) and with (b) 
a $C_{2v}$ distortion, using the SC216 supercell. The solid black (green) lines are the bands 
of the defective (perfect) supercell, and the dashed lines represent the highest occupied level. Pink and cyan balls represent Te and Cd, respectively.}
\label{fig:bandas216JTQ0}
\end{center}
\end{figure}

\begin{figure}[htbp]
\begin{center}
\includegraphics[width=.50\textwidth]{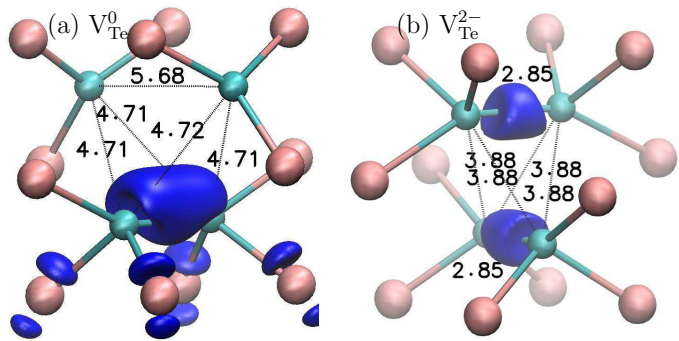}  
\caption{(Color online) Charge density isosurfaces of the defect state in the the distorted 
Te vacancy in the $0$ (a) and $2-$ (b) charge states. Pink and cyan balls represent Te and Cd, respectively. The numbers indicate the Cd-Cd distances.}
\label{fig:HOMOQ0_JT}
\end{center}
\end{figure}

The neutral state shows a similar band structure if the atomic positions keep the $T_d$ symmetry.  
However, the total energy and the band diagram are modified due to a symmetry break, 
inducing a $C_{2v}$ distortion that lowers the total energy by 0.3 eV. The distortion leads to formation of 
a Cd$_2$ dimer with a bond length of 2.86 \AA.  Figure \ref{fig:bandas216JTQ0} shows both the symmetric 
and distorted configurations, along with their band diagrams. The origin of the distortion is not due to
the Jahn-Teller mechanism, because the symmetric state does not present degeneracy. As discussed 
above for the $2-$ state, the missing Te atom leaves four broken bonds with 0.5 electron each one. 
The energy is reduced when two Cd atoms loss 0.5 electrons  and move to adopt a quasi-planar bonding structure 
with their three Te neighbors. The rules of covalent bonding are satisfied if these two Cd atoms 
distribute their 1.5 electrons in three $sp^2$ orbitals that make $\sigma$ bonds with the Te 
$sp^3$ orbitals. The other two Cd atoms, which also have $0.5\times 2=1$ electron in dangling bonds,  
trap the electron released by the first Cd pair, getting closer to form a doubly-occupied covalent bond. 
 This bond corresponds to the VB top in Fig.~\ref{fig:bandas216JTQ0}(b), where the corresponding charge density 
(square of the wave function) is shown in Fig.~\ref{fig:HOMOQ0_JT}(a).

\begin{figure}[htbp]
\begin{center}
 \includegraphics[width=.45\textwidth]{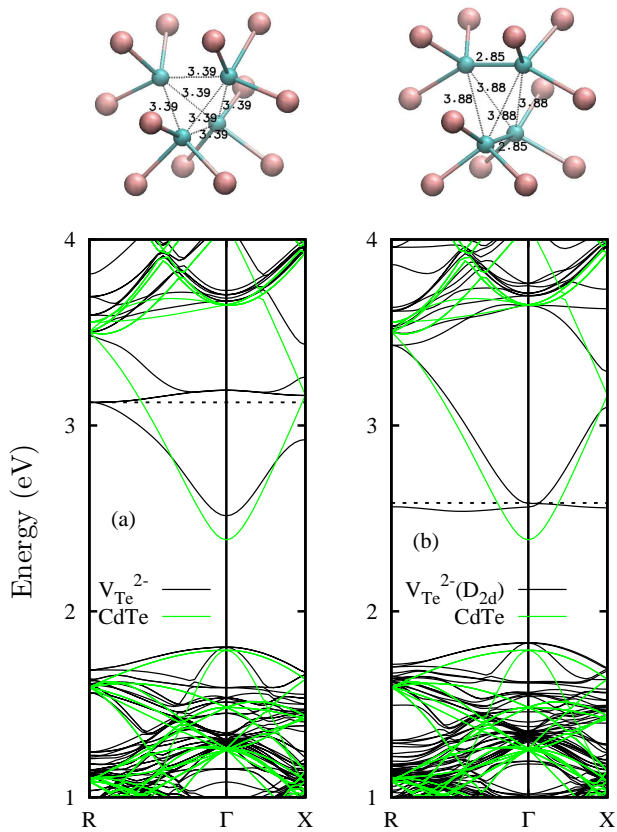}      
\caption{(Color online) Stable geometries and band diagrams for Te vacancy in the $2-$ charge state without (a), 
and with (b) a $D_{2d}$ distortion, using the SC216 supercell. The solid black (green) lines are 
the bands of the defective (perfect) supercell, and the dashed lines represent the highest occupied level. Pink and cyan balls represent Te and Cd, respectively.}
\label{fig:bandas216JTQ2-}
\end{center}
\end{figure}

For the $2-$  charge state, the total energy is apparently minimized by a $D_{2d}$ distortion leading to the  
formation of two perpendicular  Cd dimers, as illustrated in Fig. \ref{fig:bandas216JTQ2-}.
This case is instructive about the 
interplay between the different size effects.
 The band diagrams of 
Fig. \ref{fig:bandas216JTQ2-} suggests that the band energy is minimized by a downshift 
of the defect band, that becomes occupied, while the dispersive host-like conduction band gets empty.  
This mechanism lowers the total energy by 0.6~eV.  However, the self-consistent energy must be corrected 
by subtracting the band filling energy, which is -0.87~eV for the undistorted configuration, but it is 
negligible for the distorted one. Therefore, the symmetric configuration has lower energy. In order 
to test this assertion, we have inserted the distorted configuration into a larger supercell composed 
of $4\times 4\times 4$ conventional unit cells, i.e. the SC512 model, which allows 
to use only the  $\Gamma$ point. With this setup,  the band 
filling error is null. As expected, the distorted configuration in the SC512 relaxes to the non-distorted structure. 
Therefore, the distortion around the vacancy for the $2-$ state is a supercell-size effect driven by 
the conduction band filling in the SC64 and SC216 models. 
According to the above discussion, the distorted configuration is unstable within 
the PBE approximation in the limit of dilute impurity. 
However, the quasiparticle 
corrections (see Appendix \ref{app:qp}) increase the total energy of the symmetric configuration, 
leaving the distorted configuration stable.  The symmetric state has a PHS recognizable 
as the dispersive CB in 
Fig. \ref{fig:bandas216JTQ2-} (a), which is occupiedby   
 two electrons.  On the other hand, in the distorted configuration, the two extra electrons 
 occupy a localized state, shown in Fig.~\ref{fig:HOMOQ0_JT}(b), resulting in the flat band of  
Fig.~\ref{fig:bandas216JTQ2-} (b). To assess the PBE gap error, we have computed the energy levels at 
the $\Gamma$ point, using the HSE06 functional.
In this calculation, only the $\Gamma$ point was used to obtain 
the self-consistent charge density. This approximation is supported by comparing the corresponding PBE 
energies and wave functions using different k-point samplings, the $\Gamma$ point and the $2\times 2\times 2$ mesh.
 The effect of changing the  PBE functional  by the HSE06 
functional is the shift of the PHS energy by 0.4~eV. In contrast, the defect  level is shifted 
by only -0.07 eV. Additional GW calculations for the primitive cell, show that the CBM is 0.481 eV over the 
PBE value (see Appendix \ref{app:qp}). Considering that the PHS is occupied by two electrons, the total 
energy of the symmetric configuration
must be corrected by $2\times 0.481$~eV. However, for the distorted configuration the quasiparticle 
correction is quite small and can be neglected. 
As a consequence, the distorted configuration has the lowest energy. 

For the neutral state, there is no need of band filling  corrections because there is neither holes in the VB 
nor electrons in the CB. Therefore the distorted configuration is stable. The calculation with 
the model SC512 confirms this result. Moreover, the absence of conduction electrons,  
valence holes, and net charge, renders quasiparticle corrections unnecessary.

The $1+$ charge state has one electron occupying the CB in a PHS. The HSE06 calculation 
confirms that the PHS has lower energy than the localized  defect state. It seems contradictory that for the $1+$ state, with one missing electron, the highest occupied level  jumps 
into the  CB, while for the $2+$ state the highest occupied level returns to the VB. This behavior is 
due to the strong 
relaxation of the four neighbor Cd atoms into planar configurations CdTe$_3$, favoring a $sp^2$-like 
hybridization. Within a $sp^2$ shell structure, each Cd atom shares its 2 valence electrons with $3\times 1.5$ 
electrons coming from its three neighboring Te, resulting in 0.5 electron in excess of the $sp^2$ shell. 
Adding the four Cd atoms, this sums 2 electrons. The net effect is that two  electronic levels are removed 
(holes). Therefore, the closed-shell structure is recovered in the charge state $2+$,  while for the state $1+$ 
there is one electron in the CB. Accordingly, for the $1+$ charge state, the formation energy must be corrected due to errors in both the 
conduction band filling and the VB and CB edges. For the latter, $z_e=1$ and $z_h=2$ must be considered 
in Eq.~\ref{gapcorr},  respectively.

\subsection{Formation energy}

\begin{figure}[htbp]
\begin{center}
\includegraphics[width=8.5cm]{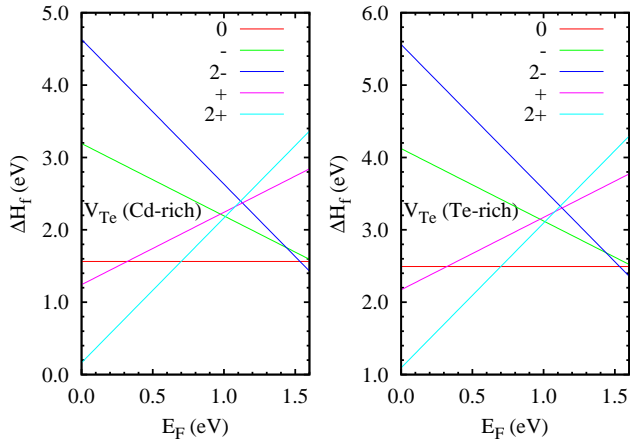}
\caption{(Color online) Formation energies ($\Delta$H) as a function of the Fermi energy (E$_{\rm F}$) 
computed with a 64-atom supercell (SC64), not considering distortions.}
\label{fig:VTe64}
\end{center}
\end{figure}

Figure \ref{fig:VTe64} 
shows the formation energies of the \defe{V}{Te} defect performed with the SC64 supercell in several 
charge states. These energies do not contain the band filling and quasiparticle corrections. However, 
these corrections partially cancel each other,  providing reasonable results. Thus, 
the differences with previous studies \cite{zhangwei2002,dusingh08,du2012} can be easily understood. 
In Fig.~\ref{fig:VTe64}, the negatively charged states appear stable for high values of the Fermi level 
because the energies of the excess electrons in the CB is heavily underestimated and, as discussed 
above, the quasiparticle correction (Eq.~\ref{gapcorr}) could not be applied.  Recently, Biswas and 
Du\cite{du2012} using the hybrid PBE0 functional tuned to fit the CdTe band gap, found no stable 
negative state. Well before, Wei and Zhang\cite{zhangwei2002} using LDA, applied several approaches 
 that led to a large band gap and no negative stable states.  Later on,
Du et al\cite{dusingh08}, using the LDA with quasiparticle correction, found that the stable states are 
$2+, 0$, and $2-$. Their results were attributed to $C_{2v}$ and $D_{2d}$ distortions that reduce 
the energies of the states $0$ and $2-$, respectively. 
The formation energies of Fig.~\ref{fig:VTe64} correspond to undistorted defects, obtained by relaxation 
of the structure after removal of a Te atom. Indeed, we found distorted configurations that lower the 
energy of the $0$ and $2-$ states.  According to the discussion of Section \ref{sec:iiia}, we will analyze the distorted structures only for the larger SC216 supercell. 

Recent calculations using the HSE06 method\cite{weiprl2013,wei2014} suggest that  \defe{V}{Te} is stable 
only in the  $2+$ charge state. These authors attribute the high stability of the charged state to the large Cd 
displacements around the vacancy. We think these authors may have missed the energy downshift of the 
neutral and $2-$ 
states due to distortion. In any case, we consider unreliable the results obtained with the SC64 model.

The formation energies, computed with the SC216 model, are given as
\begin{subequations}
\begin{eqnarray}
\Delta  H_f ({\V_{\Te}}^0 ) &=&2.455\mbox{ eV} + \Delta \mu_{Te} \\
\Delta  H_f ({\V_{\Te}}^{1-} )&=&4.532\mbox{ eV} + \Delta \mu_{Te} - E_F \\
\Delta  H_f ({\V_{\Te}}^{2-} )&=&5.845\mbox{ eV} + \Delta \mu_{Te} -2E_F \\
\Delta  H_f ({\V_{\Te}}^{1+} )&=&2.371\mbox{ eV} +\Delta \mu_{Te} +E_F\\
\Delta  H_f ({\V_{\Te}}^{2+} )&=&1.072\mbox{ eV} + \Delta \mu_{Te} +2E_F
\end{eqnarray}
\end{subequations}

\begin{figure}[htbp]
\begin{center}
\includegraphics[width=8.5cm]{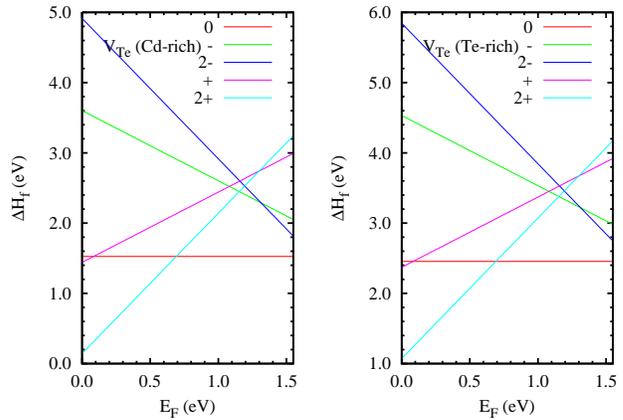}
\caption{(Color online) Formation energies as a function of the Fermi energy for the Te vacancy computed with 
a 216-atom supercell (SC216).}
\label{fig:VTe256}
\end{center}
\end{figure}

Figure \ref{fig:VTe256} 
shows the formation energy  plots as function of the Fermi level. 
The quasiparticle corrections has been 
applied according to the Appendix \ref{app:qp}, shifting the VBM by -0.470~eV and the 
CBM by +0.481~eV.  Remarkably, the formation energy of the stable charge states $0$ and $2+$ are 
insensitive  to the quasiparticle corrections. These corrections are not required for the neutral state, while for the 
$2+$ state the terms (\ref{gapcorr}) and (\ref{gapcorr2}) cancel each other. The net quasiparticle correction
for the states $2-$, $1-$, and $1+$ are 0.939, 0.951 and 0.951~eV, respectively. 
As this method gives a band gap of 1.55~eV, 
this is the maximum value for the Fermi level in Fig.~\ref{fig:VTe256}. The thermodynamic 
transition level $\varepsilon(2+/0)=0.69$ eV. It is worth to note that almost the same value for  
$\varepsilon(2+/0)$ is obtained with the model SC64 without band-filling and gap corrections 
 (see Fig.~7). For the other charge states, the observed good agreement between Figs.~\ref{fig:VTe64} and \ref{fig:VTe256} 
  is due to error compensations between the gap error and the band-filling error.  

Some calculations have predicted negative formation energies for 
\defe{V}{Te}, mostly in Te-poor conditions\cite{du2012,weiprl2013}, but also 
in Te-rich conditions\cite{wei2014}. Negative formation energies indicate a trend to high 
defect concentration
and agglomerate. High defect concentration need to be studied by a different methodology
 that correct this apparently wrong result. However, high concentration of \defe{V}{Te} in CdTe is 
not observed\cite[Ch.\ 5]{librocdte}. 
It is worth to note that our calculations always predict positive formation energies for \defe{V}{Te}.

\section{Conclusions}

In summary,
the use of the 64-atoms supercell, as the basis to calculate defects in CdTe and other semiconductors, 
is revisited. For the Te vacancy, \defe{V}{Te}, the band structure of 
some charge states 
presents unphysical features that are easily corrected by using larger supercells. These unphysical features, 
notably a dramatic alteration of the conduction band, makes uncertain to apply simple quasiparticle corrections 
to the band edges. The formation energies and thermodynamic transition levels are slightly modified by these  errors 
 due to the partial compensation of the band-filling error and the gap underestimation 
of GGA functionals.  
  
The band-filling error associated to the k-point sampling  can drive the system to  distorted configurations, 
as the case of \defeq{V}{Te}{2-}. The calculations based on the 512-atoms cubic supercell SC512, with 
$\Gamma-$point sampling, are free of this error. However, the 
symmetric configuration turns out to have higher energy after quasiparticle (gap) corrections. In this sense, the band-filling error mimics the quasiparticle corrections, allowing to relax the system to the distorted configurations.  
For calculations when the SC512 
model is not affordable, we suggest a careful analysis of the band-filling errors in both distorted and undistorted 
configurations. High level calculations like 
hybrid functionals, GW, or quantum Monte Carlo, are probably not affordable for such large supercells, and may need to be 
done for smaller supercells. In these cases, further optimization of the distorted structures inserted 
in a large 
supercell with a single k-point and a low-cost GGA functional, are recommended. 
 
 According to the literature, the
Te vacancy seems to be the dominating donor and has strong influence on the electric and thermodynamical properties of 
CdTe\cite{du2012,weiprl2013,wei2014}. Our study shows that predictions based on calculations with the  SC64 model are 
questionable, and  additional efforts must be made, considering larger supercells like SC216 or SC512
 in order to obtain reliable results.
 
\acknowledgments
This work was supported by the FONDECYT Grant No. 1130437.
The authors thankfully acknowledge the computer resources, technical expertise and assistance provided by the 
Madrid Supercomputing and Visualization Center (CeSViMa) and Red Espa\~nola de Supercomputaci\'on, where part 
of these calculations were done, and Prof. P. Wahn\'on for her hospitality and useful discussions at Universidad Polit\'ecnica de Madrid. 
WO acknowledges support from CONICYT-PIA under the Grant Anillo ACT-1107.  

\appendix
\section{Size corrections}
\label{app:size}
An isolated defect either populates the conduction band minimum (CBM) or create holes at the valence band 
maximum (VBM). However, in a finite supercell, when a k-point mesh is used, the electron or hole created by 
the point defect fills half a band, considering the spin degeneracy. The single-particle contribution to the 
total energy is affected by the band dispersion. The correction is to replace the energies of the defect band, 
either of electrons or holes, by the corresponding band extremum. For donors, the correction can be expressed 
as\cite{lanyzunger08prb}
\begin{equation}
 \Delta E_{b.f.} = - \sum_{n,k} \Theta(e_{n,k}-e_{min})w_k \eta_{n,k}(e_{n,k} - e_{min}),
\end{equation}
where indexes $n$ and $k$ denote the bands and k-points, $w_k$ are the k-point weights,  $e_{n,k}$ are the single particle energies with 
minimum $e_{min}$, $\eta_{n,k}$ are the occupation numbers,  and $\Theta$ is the Heaviside function. The correction for acceptors is 
\begin{equation}
 \Delta E_{b.f.} =  \sum_{n,k} \Theta(e_{max}-e_{n,k}) w_k (S-\eta_{n,k})(e_{max}-e_{n,k} ), 
\end{equation}
where $S=1$ ($S=2$) stands for spin polarized (unpolarized) calculations. $e_{max}$ is the maximum of the acceptor 
band. 

In VASP calculations, the average of the electrostatic potential is set to zero. Hence, the 
potential energy reference is undetermined by an additive constant, which is different in supercells containing 
different number of atoms or different charge. The change has two sources: a) the presence of excess charge, 
b) the missing or appearance of extra atoms. For charged systems, the additive constant enters in the Hartree 
term of the total energy per cell:
$$
E_H= \int e [n_0(\mathbf{r})- N(\mathbf{r})+\Delta n(\mathbf{r} )](v(\mathbf{r}) + \Delta v) d^3 {\mathbf{r}} ,
$$
where $e$ is the electron charge, $n_0(\mathbf{r})$ is the electron density for a neutral system, 
$\Delta n(\mathbf{r})$ is the excess electron density. $N(\mathbf{r})$ is the nuclear charge, and 
$\Delta v$ is the  additive potential implicitly added to make zero the average potential, with respect to 
the potential of the system without defect and neutral. The contribution of the additive potential to the Hartree 
energy is equal to
\begin{eqnarray*}
\Delta E_H &=& \int e \Delta n(\mathbf{r})\Delta v d^3 {\mathbf r} \\ 
&=& \left[ \int \Delta n(\mathbf{r}) 
d^3 {\mathbf r} \right] (e \Delta v ) 
 =  (-q) \Delta V.
\end{eqnarray*}
Here, $q$ is the number of electrons donated $\int \Delta n(\mathbf{r} d^3 {\mathbf r} )/(-e)$, and 
$\Delta V=e\Delta v$ is the change of electron potential energy. $\Delta V$ can be estimated by the core level 
shifts of atoms far from the defect with respect to the core levels in the calculation with the perfect crystal. 
The correction to the total energy due to potential alignment (p.a.) is 
\begin{equation}
\Delta E_{p.a.}=q\Delta V.
\end{equation}

Periodic boundary conditions induce an spurious interactions of the carved defect with the periodic replicas. 
The leading terms can be corrected by the Makov-Payne formula
$$
\Delta E_{M-P}=\frac{\alpha_M q^2}{2 \epsilon V^{1/3}} + \frac{2 \pi q Q }{3 \epsilon V} + O(V^{-5/3}),
$$
where $\alpha_M$ is a Madelung constant, $\epsilon$ is the dielectric constant, $V$ is the supercell volume, 
and $Q$ is a quadrupole moment. We have used a simplified version due to Lany and Zunger\cite{lanyzunger08prb} 
\begin{equation}
\Delta E_{L-Z}=\frac{2}{3}\frac{\alpha_M q^2}{2 \epsilon V^{1/3}} .
\end{equation}
We have used the static dielectric constant $\epsilon=12.272$, obtained from density functional perturbation theory calculations and including local field effects\cite{dielectricinvasp,baronidfpt}. 

In summary, the supercell size corrections are given by the sum of band-filling, potential alignment, and Lany-Zunger corrections
\begin{equation}
\Delta E_{size}=\Delta E_{b.f.} + \Delta E_{p.a.} + \Delta E_{L-Z}. 
\end{equation}

\section{Quasiparticle corrections}
\label{app:qp}
Semilocal exchange-correlation functionals (LDA, GGA) cause a severe underestimation of the valence to conduction 
band gap, and this affects the formation energies. Let us consider the gap error
$$
\Delta E_g = E_g^{exp} - E_{gap}^{DFT} = \Delta E_C - \Delta E_V.
$$
The above equation established a partition of the gap error in shifts applied to the valence and conduction band. 

For shallow acceptors, where active holes occupy perturbed host states, the hole levels undergo the same shift as the valence band edge. 
This has been verified in quasiparticle calculations\cite{lanyzunger08prb}. For shallow donors, the active electron levels are perturbed 
states of the conduction band, undergoing the shift. Including fractional occupations, the shifts of single particle 
levels and total energies can be expressed as 
\begin{equation}
\label{gapcorr}
\Delta E_g^{(1)}(q)= \left\{ 
\begin{array}{cc}
 z_e(X,q)\Delta E_C & \mbox{ for donors,}  \\
 z_h(X,q) (-\Delta E_V) & \mbox{ for acceptors,} 
\end{array} \right.
\end{equation}
where $z_e$ ($z_h$) is the number of electrons introduced (removed) by the defect near the CBM (VBM).

A second quasiparticle correction modifies the reference of the Fermi level
\begin{equation}
\label{gapcorr2}
\Delta E_g^{(2)}(q) = q \Delta E_V .
\end{equation}
The total quasiparticle correction is 
\begin{equation}
\Delta E_{q.p.}= \Delta E_g^{(1)}(q)+\Delta E_g^{(2)}(q) .
\end{equation}
 
 With our computational setup with the PBE functional, the VBM and CBM energies are 1.789 and 2.388 eV, respectively. For quasiparticle corrections we have considered the generalized DFT with the hybrid functional HSE06\cite{hse06} and GW calculations\cite{shishkin07c}, including the spin-orbit coupling, 
 for the CdTe primitive cell, using a $8\times 8\times 8$ k-point grid. 
 The HSE06
  VBM and CBM energies are 1.773 and 2.977 eV, respectively. With this approximation the band gap is 1.204 eV, which is still lower than the experimental value.
The single-shot perturbative G$_0$W$_0$ calculation\cite{shishkin07c}  based upon the HSE06  calculation provides a band gap of 1.55 eV, and downshift of the VBM by 0.454 eV below the HSE06
  value. The latter HSE06 
 and G$_0$W$_0$ were performed with PAW potentials adapted to GW calculations. The used cutoff for the response function was 169.4 eV. The
  integrals defining the self-energies were evaluated using 96 frequencies between 0 and 288 eV$/\hbar$.   The  number of (occupied plus empty) bands was carried up to 1600, and the VBM was obtained by extrapolation to infinite bands.
  

\begin{thebibliography}{10}

\bibitem{shockley}
W.~Shockley and H.~J. Queisser.
\newblock Detailed balance limit of efficiency of p-n junction solar cells.
\newblock {\em J. Appl. Phys.}, 32:510, 1961.

\bibitem{librocdte}
Robert Triboulet and Paul Siffert, editors.
\newblock {\em {CdTe} and Related Compounds; Physics, Defects, Hetero- and
  Nano-structures, Crystal Growth, Surfaces and Applications}.
\newblock European Materials Research Society Series. Elsevier, Amsterdam,
  2010.

\bibitem{vte_epr}
B.~K. Meyer, P.~Omling, E.~Weigel, and G.~M\"uller-Vogt.
\newblock F center in {CdTe}.
\newblock {\em Phys. Rev. B}, 46:15135, 1992.

\bibitem{lany2001vte}
S.~Lany, V.~Ostheimer, H.~Wolf, and Th. Wichert.
\newblock Vacancies in {CdTe}: experiment and theory.
\newblock {\em Physica B: Condensed Matter}, 308 - 310:958 -- 962, 2001.

\bibitem{vcd_hse}
Xu~Run, Hai-Tao Xu, Min-Yan Tang, and Lin-Jun Wang.
\newblock Hybrid density functional studies of cadmium vacancy in {CdTe}.
\newblock {\em Chin. Phys. B}, 23:077103, 2014.

\bibitem{illgner96vte}
M.~Illgner and H.~Overhof.
\newblock Electronic structure and hyperfine interactions for deep donors and
  vacancies in ii-vi compound semiconductors.
\newblock {\em Phys. Rev. B}, 54:2505--2511, 1996.

\bibitem{verstraten2003}
D.~Verstraeten, C.~Longeaud, A.~Ben~Mahmoud, H.~J. von Bardeleben, J.~C.
  Launay, O.~Viraphong, and Ph.~C. Lemaire.
\newblock A combined {EPR} and modulated photocurrent study of native defects
  in {Bridgman} grown vanadium doped cadmium telluride: the case of the
  tellurium antisite.
\newblock {\em Semicond. Sci. Technol.}, 18:919, 2003.

\bibitem{dusingh08}
Mao-Hua Du, Hiroyuki Takenaka, and David~J. Singh.
\newblock Carrier compensation in semi-insulating {CdTe}: First-principles
  calculations.
\newblock {\em Phys. Rev. B}, 77:094122, 2008.

\bibitem{du2012}
K.~Biswas and Mao-Hua Du.
\newblock What causes high resistivity in {CdTe}.
\newblock {\em New J. Phys.}, 14:063020, 2012.

\bibitem{weiprl2013}
Jie Ma, Darius Kuciauskas, David Albin, Raghu Bhattacharya, Matthew Reese,
  Teresa Barnes, Jian~V. Li, Timothy Gessert, and Su-Huai Wei.
\newblock Dependence of the minority-carrier lifetime on the stoichiometry of
  {CdTe} using time-resolved photoluminescence and first-principles
  calculations.
\newblock {\em Phys. Rev. Lett.}, 111:067402, 2013.

\bibitem{wei2014}
Ji-Hui Yang, Ji-Sang Park, Joongoo Kang, Wyatt Metzger, Teresa Barnes, and
  Su-Huai Wei.
\newblock Tuning the {Fermi} level beyond the equilibrium doping limit through
  quenching: The case of {CdTe}.
\newblock {\em Phys. Rev. B}, 90:245202, 2014.

\bibitem{zhangwei2002}
Su-Huai Wei and S.~B. Zhang.
\newblock Chemical trends of defect formation and doping limit in ii-vi
  semiconductors: The case of {CdTe}.
\newblock {\em Phys. Rev. B}, 66:155211, 2002.

\bibitem{cdte_ggau}
E.~Men\'endez-Proupin, A.~Am\'ezaga, and N.~Cruz~Hern\'andez.
\newblock Electronic structure of {CdTe} using {GGA+U}$^{\mathrm{sic}}$.
\newblock {\em Physica B}, 452:119 -- 123, 2014.

\bibitem{paw1}
P.~E. Bl\"ochl.
\newblock Projector augmented-wave method.
\newblock {\em Phys. Rev. B}, 50:17953, 1994.

\bibitem{paw2}
G.~Kresse and D.~Joubert.
\newblock From ultrasoft pseudopotentials to the projector augmented wave
  method.
\newblock {\em Phys. Rev. B}, 59:1758, 1999.

\bibitem{vasp4}
G.~Kresse and J.~Furthm{\"u}ller.
\newblock Efficient iterative schemes for ab initio total-energy calculations
  using a plane-wave basis set.
\newblock {\em Phys. Rev. B}, 54:11169--11186, 1996.

\bibitem{pbe}
John~P. Perdew, Kieron Burke, and Matthias Ernzerhof.
\newblock Generalized gradient approximation made simple.
\newblock {\em Phys. Rev. Lett.}, 77:3865--3868, 1996.

\bibitem{hse06}
J.~Heyd, G.~E. Scuseria, and M.~Ernzerhof.
\newblock Erratum: ``{H}ybrid functionals based on a screened {C}oulomb
  potential´´ [{J. Chem. Phys.} 118, 8207 (2003).
\newblock {\em J. Chem. Phys.}, 124:219906, 2006.

\bibitem{medea}
MedeA$^{\mbox{\textregistered}}$ Materials Design, Inc., Santa Fe, New Mexico,
  USA, 2013.

\bibitem{PHS}
Stephan Lany and Alex Zunger.
\newblock Anion vacancies as a source of persistent photoconductivity in
  {II-VI} and chalcopyrite semiconductors.
\newblock {\em Phys. Rev. B}, 72:035215, 2005.

\bibitem{lanyzunger08prb}
Stephan Lany and Alex Zunger.
\newblock Assessment of correction methods for the band-gap problem and for
  finite-size effects in supercell defect calculations: Case studies for {ZnO}
  and {GaAs}.
\newblock {\em Phys. Rev. B}, 78:235104, 2008.

\bibitem{dielectricinvasp}
Xifan Wu, David Vanderbilt, and D.~R. Hamann.
\newblock Systematic treatment of displacements, strains, and electric fields
  in density-functional perturbation theory.
\newblock {\em Phys. Rev. B}, 72:035105, 2005.

\bibitem{baronidfpt}
Stefano Baroni, Stefano de~Gironcoli, Andrea Dal~Corso, and Paolo Giannozzi.
\newblock Phonons and related crystal properties from density-functional
  perturbation theory.
\newblock {\em Rev. Mod. Phys.}, 73(2):515--562, Jul 2001.

\bibitem{shishkin07c}
F.~Fuchs, J.~Furthm\"uller, F.~Bechstedt, M.~Shishkin, and G.~Kresse.
\newblock Quasiparticle band structure based on a generalized {K}ohn-{S}ham
  scheme.
\newblock {\em Phys. Rev. B}, 76, 2007.

\end{thebibliography}

\end{document}